\documentclass[12pt,preprint]{aastex}

\slugcomment{To appear in ApJ GALEX Special Issue}

\shorttitle{UV Spectroscopy of LIRGs in ELAIS S1}
\shortauthors{Burgarella et al.}

\begin{document}


\title{GALEX Ultraviolet Spectroscopy and Deep Imaging of Luminous InfraRed Galaxies in the ELAIS S1 field}

\author{Denis Burgarella\altaffilmark{1}, V\'eronique Buat\altaffilmark{1},
Todd Small\altaffilmark{2}, Tom A. Barlow\altaffilmark{2},
Samuel Boissier\altaffilmark{3}, 
Armando Gil de Paz\altaffilmark{3}, Timothy M. Heckman\altaffilmark{4},
Barry F. Madore\altaffilmark{3,5}, D. Christopher Martin\altaffilmark{2},
R. Michael Rich\altaffilmark{6}, Luciana Bianchi\altaffilmark{7},
Yong-Ik Byun\altaffilmark{8},
Jose Donas\altaffilmark{1}, Karl Forster\altaffilmark{2},
Peter G. Friedman \altaffilmark{2}, Patrick N. Jelinsky\altaffilmark{6}, 
Young-Wook Lee\altaffilmark{5}, Roger F. Malina\altaffilmark{1},
Bruno Milliard\altaffilmark{1}, Patrick Morrissey \altaffilmark{2},
Susan G. Neff\altaffilmark{7}, David Schiminovich \altaffilmark{2},
Oswald H.W. Siegmund\altaffilmark{6}, Alex S. Szalay\altaffilmark{4},
Barry Y. Welsh\altaffilmark{6} \and Ted K. Wyder\altaffilmark{2}}

\affil{Observatoire Astronomique Marseille Provence, Laboratoire d'Astrophysique de Marseille, traverse du siphon, 13376 Marseille cedex 12, France; denis.burgarella@;veronique.buat@;jose.donas@;roger.malina@;bruno.milliard@oamp.fr}
\affil{California Institute of Technology, MC 405-47, 1200 E. California Boulevard, Pasadena, CA 91125, USA; tas@;tab@;cmartin;krl@;friedman@;patrick@;ds@;wyder@srl.caltech.edu}
\affil{Observatories of the Carnegie Institution of Washington, 813 Santa Barbara Street, Pasadena 91101, USA; boissier@ociw.edu,agpaz@;barry@ipac.caltech.edu}
\affil{Department of Physics and Astronomy, Johns Hopkins University, 3400 North Charles Street, Baltimore, MD 21218; heckman@adcam.pha.jhu.edu;szalay@tardis.pha.jhu.edu}
\affil{NASA/IPAC Extragalactic Database, California Institute of Technology, Mail Code 100-22, 770 S. Wilson Ave., Pasadena, CA 91125, USA}
\affil{Department of Physics and Astronomy, University of California, Los Angeles, CA 90065, USA; rmr@astro.ucla.edu}
\affil{Center for Astrophysical Sciences, The Johns Hopkins University, 3400 North Charles St., Baltimore, MD 21218, USA; bianchi@skyrv.pha.jhu.edu}
\affil{Center for Space Astrophysics, Yonsei University, Seoul 120-749, Korea; byun@;ywlee@csa.yonsei.ac.kr}
\affil{Space Sciences Laboratory, University of California at Berkeley, 601 Campbell Hall, Berkeley, CA 94720, 
USA; patj@;ossy@;bwelsh@ssl.berkely.edu}
\affil{Laboratory for Astronomy and Solar Physics, NASA Goddard Space Flight Center, Greenbelt, MD 20771, USA; neff@cobblr.gsfc.nasa.gov}


\begin{abstract}
The ELAIS S1 field was observed by GALEX in both its Wide Spectroscopic and Deep Imaging Survey modes. This field was previously observed by the Infrared Space Observatory and we made use of the catalogue of multi-wavelength data published by the ELAIS consortium to select galaxies common to the two samples. Among the 959 objects with GALEX spectroscopy, 88 are present in the ELAIS catalog and 19 are galaxies with an optical spectroscopic redshift. The distribution of redshifts covers the range $0<z<1.6$. The selected galaxies have bolometric IR luminosities $10<Log(L_{IR})<13$ (deduced from the $15 \mu m$ flux using ISOCAM) which means that we cover a wide range of galaxies from normal to Ultra Luminous IR Galaxies. The mean ($\sigma$) UV luminosity (not corrected for extinction) amounts to $Log(\lambda.L_{1530}) = 9.8 (0.6)$ $L_\sun$ for the low-z ($z \le 0.35$) sample. The UV slope $\beta$ (assuming $f_\lambda \propto \lambda^\beta$) correlates with the GALEX FUV-NUV color if the sample is restricted to galaxies below $z < 0.1$. Taking advantage of the UV and IR data, we estimate the dust attenuation from the IR/UV ratio and compare it to the UV slope $\beta$. We find that it is not possible to uniquely estimate the dust attenuation from $\beta$ for our sample of galaxies. These galaxies are highly extinguished with a median value $A_{FUV} = 2.7 \pm 0.8$. Once the dust correction applied, the UV- and IR-based SFRs correlate. For the closest galaxy with the best quality spectrum, we see a feature consistent with being produced by a bump near 220nm in the attenuation curve.
\end{abstract}



\clearpage


\section{Introduction}

GALEX, the Galaxy Evolution Explorer is a $50-cm$ telescope devoted to observing the Ultraviolet (UV in two bands centered at about 153m and 231nm) sky in several Imaging Surveys including an All-sky Imaging Survey and several Spectroscopic Surveys (Martin et al. 2004, Barlow et al. 2004). This paper presents the first analysis of the GALEX Early Release spectroscopic Observations available from GALEX. The European Large Area ISO Survey (ELAIS) is a database of deep $15\mu m$ observation carried out with ISO to study the extragalactic universe in the IR range. The correlation of ELAIS with UV data allows us to carry out a multi-wavelength study and to estimate the dust attenuation to deduce and compare star formation rates (SFR) estimated from UV and IR. Applications of this work are related to Lyman Break Galaxies discovered at $z=3$ (Steidel et al. 1996, Lowenthal et al. 1997) and now to even higher redshift galaxies at $z \approx 6$ with the Hubble Ultra Deep Field by Bunker et al. (2004) and maybe $z=10$ (Pello et al. 2004). Although the present sample in not purely UV-selected, it aims at better understanding how to correct for the dust attenuation in these galaxies. It could also lead to a way of understanding UV characteristics of IR-bright galaxies.

We assume a cosmology with $H_0 = 70 km.s^{-1}.Mpc^{-1}$, $\Omega_M = 0.3$ and $\Omega_{VAC}=0.7$ in this paper.

\section{The Galaxy Sample}


GALEX observed the ELAIS S1 field for 12198 sec. in the imaging mode on 15 September 2003 and 31267 sec. in the spectroscopic mode on 24 September 2003. This field had already been observed by the Infrared Space Observatory (ISO) using ISOCAM. The southern field, S1, centered at $\alpha$ = 00h34m444, $\delta$ = -43deg28'12'' (J2000.0), covers an area of the sky of about 2 $\times$ 2 deg$^2$. The final band-merged ELAIS catalogue (Rowan-Robinson et al. 2003), containing 3523 sources, is available through the ELAIS web site. The S1 catalog (complete at the 5 $\sigma$ level) contains 462 sources in the flux density range 0.5-150 mJy (La Franca et al. 2004). GALEX IR0.2 pipeline detected 959 spectra within the 1.2-deg field of view that we cross-correlated with the ELAIS S1. The area of the common field is about 0.5 deg$^2$. Our ELAIS S1 sample was selected by cross-correlating the UV GALEX spectroscopic sample with the Mid IR (MIR at $15 \mu m$) ISO catalogue assuming that the emission from objects within a circular aperture of 6-arcsec radius could be from the same object. This cross-correlation provided us with 88 objects, most of which proved to be galactic foreground stars. From this sample, 19 galaxies with spectroscopic redshifts were finally selected (Table 1) and we focus on this sample to carry out a multi-wavelength analysis. All the objects except two are below $z = 0.35$ : their average FUV luminosity is $Log(L_{FUV}) = 9.8 (\pm 0.6)$ $L_\sun$ and their average IR luminosity $Log(L_{IR}) = 11.1 (\pm 0.5)$ $L_\sun$ where Log(L) are luminosities in erg.s$^{-1}$. Assuming Luminous IR Galaxies (LIRGs) luminosities in the range $10^{11} \leq L_{IR}/L_\sun < 10^{12}$ and Ultra Luminous IR Galaxies (ULIRGs) have $L_{IR}/L_\sun > 10^{12}$ (Chary \& Elbaz 2001, Goldader et al. 2002), 12 of them qualify as LIRGs, 4 objects are Seyfert 1 and 1 object is classified as a Liner. Two objects are at $z > 1$, their UV luminosities are above $Log(L_{UV}) = 12$ and their IR luminosities are about $Log(L_{IR}) = 13$ i.e. Ultra Luminous IR Galaxies (ULIRGs). However, since they are classified by ELAIS as Seyfert 1, the contribution of the central non-thermal QSO is very likely to be predominent for these objects and their study is outside the scope of the present paper which is to study star-forming objects and their dust attenuation.

\section{GALEX Ultraviolet Colors and the UV Slope $\beta$}

The interpretation of UV spectra must mainly deal with two parameters : the stellar population (with associated parameters : star formation rate, age, metallicity, etc.) and the dust reddening (with associated parameters : attenuation curve, amount of dust, etc.). In order to get back to the former, we have to correct for the latter. To estimate this correction, Calzetti, Kinney \& Storchi-Bergmann (1994) fitted the observed UV continuum (corrected for the foreground extinction) by a powerlaw expressed as $f_\lambda \propto \lambda^\beta$ in the range $1250 \leq \lambda (\AA) \leq 2600$. They selected windows to exclude absorption lines and the $2175 \AA$ dust feature which might cause a departure from the assumed power law. For their sample of starbursts, the resulting $\beta$ was found to correlate with the Balmer optical depth and therefore with the dust content. One interesting point is that the attenuation curve deduced from Calzetti, Kinney \& Storchi-Bergmann's (1994) sample does not show any evidence from the $2175 \AA$ dust feature.

We estimated the UV slope $\beta$ of our sample after visually inspecting the spectra and discarding pixels in absorption lines or where the signal-to-noise ($S/N$) was low. Following Calzetti, Kinney \& Storchi-Bergmann, we assumed that no $2175 \AA$ dust feature is present and keep the corresponding pixels in the fit. The slopes obtained from the fit over the left pixels are listed in Table 1 with the associated uncertainty $\sigma_\beta$. Figure~1 presents one of the best and one of the worst fits over the observed spectra. For ELAISC15\_J0035-4356, lying at the edge of the field of view and close to a bright object, the slope cannot be estimated. For most of the GALEX database, we have the UV color $FUV-NUV$ which can be used (Buat et al. 2004, Schiminovich et al. 04, Arnouts et al. 2004) to evaluate the UV slope $\beta$. So, it is interesting to compare the estimated $\beta$ to the $FUV-NUV$ colors (Figure 2). There is no apparent relationship between the GALEX $FUV-NUV$ color and the UV slope $\beta$ for the complete sample. However, as shown by the analysis of FUSE spectra by Buat et al. (2002) and expected from models (e.g. Leitherer et al. 1999), the continuum of starburst galaxies flattens below $\sim 1200 \AA$ and it cannot be fitted over the full $\lambda \lambda 912\AA - 2400\AA$ by a single power law. If we restrict our sample to galaxies with $z \leq 0.1$, $\beta$ correlates with $FUV-NUV$ and the trend is very consistent with the GALEX $FUV-NUV$ vs. $\beta_{GLX}$ relationship found by Kong et al. (2004) within an absolute uncertainty of about $0.10-0.15$ mag. To further check this, we have integrated Kinney et al.'s (1993) templates into GALEX filters (except the bulge and elliptical templates). The location and standard deviation of these points is consistent with our data for $z < 0.1$. For this reason, the wavelength range used for the fit is restricted to rest-frame wavelengths $\lambda > 1200 \AA$. We suggest that any estimation of $\beta$ from $FUV-NUV$ should include a K-correction above about $z \sim 0.1$ as performed in Schiminovich et al. (2004) and Arnouts et al. (2004).

\section{Dust Attenuation and Star Formation Rates}

By construction, all our objects are detected in UV and in MIR. From the $15 \mu m$ fluxes, we evaluated the IR bolometric luminosity $L_{IR} = L (8-1000 \mu m)$ (Chary \& Elbaz 2001) and we estimated the $L_{IR}/L_{FUV}$ ratio.  The 1-$\sigma$ uncertainty for the $15 \mu m$ flux to IR bolometric luminosity relation is about 0.2 in $Log(L_{IR})$. However, by accounting for measurement errors, it could rise to 0.5 - 0.7 in $Log(L_{IR})$.

GALEX spectroscopic observations present the first opportunity of comparing actual spectroscopic $\beta$ values to the $L_{IR}/L_{FUV}$ ratio for LIRGs. This is important because estimates obtained from colors (see for instance Meurer et al. (1999) for a discussion of spectroscopic and photometric $\beta$'s) could lead to larger uncertainties in the evaluation of $\beta$ since the whole spectral energy distribution is binned to two points (with their associated uncertainties). Therefore, we loose the information about lines, the shape of the attenuation curve and the star formation history necessary to get the best $\beta$. Moreover, we do not need to correct for band-shifting and K-corrections (cf. Sect. 3).

The location of our galaxy sample in the $L_{IR}/L_{FUV}$ vs. $\beta$ diagram, plotted in Figure 3, shows that a simple law cannot represent the whole sample of galaxies. Indeed, if some of them follow Kong et al.'s (2004) IR/FUV vs. $\beta$ law for starbursts, most of them fall above the relation and suggest a higher extinction that would be evaluated from $\beta$. ULIRGs from Goldader et al. (2002) are also overplotted in this diagram. All but one lie in the top-right zone of the diagram (greater UV attenuations and flatter slopes), these objects are evidently much more extreme than ours but they seems to have a similar behavior. It is also interesting to notice that most of the Seyfert 2 galaxies have the highest $L_{IR}/L_{FUV}$, which might mean that there is an IR contribution from a dust torus around the QSO (Rowan-Robinson et al. 2003). By using the relation given by Buat et al. (2004), we can estimate the FUV dust attenuation listed in Table 1. The median value for our sample is $A_{FUV} = 2.7 \pm 0.8$ in agreement with Buat et al. (2004) median value of $A_{FUV} = 2.9^{+1.3}_{-1.1}$ for their FIR selected sample while their median $A_{FUV} = 1.1^{+0.5}_{-0.4}$ for their NUV selected sample.

From the spectrum of the nearest galaxy in the sample : ELAISC15\_J003828-433848 at $z = 0.048$, we can see a trough in the spectrum which is consistent with being produced by a bump in the attenuation law at the same location that the one present in the Milky Way attenuation curve near 220 nm (Figure 4). Within the limit of this early calibration of GALEX spectroscopy, the fit to the continuum gives $\beta = -1.58 \pm 0.04$. Assuming the presence of the 220nm bump in this LIRG, we can estimate a new slope by fitting a power-law without accounting for wavelengths in the range $200-250$ nm, i.e. where the effect of the bump is maximal (e.g. Witt \& Gordon 2000). The new value of the slope is $\beta = -1.26$ i.e. at 8 $\sigma$ from the previous value. This galaxy falls above Kong et al. law (see Figure 3) but if we use the new value of $\beta$, the galaxy shifts to the right (shallower $\beta$) and gets closer to the curve. Of course, it is difficult to draw conclusions from this single instance, but Motta et al. (2002) suggest the presence of the same feature in the spectrum of a galaxy at $z \sim 0.83$. Moreover, even though the sample is built from normal galaxies and not LIRGs, it is worth noticing that Rowan-Robinson (2003) found that a normal Galactic extintion law seems consistent with their analysis of their data. Finally, the only high $S/N$ large sample of UV spectra is Kinney et al. (1993) atlas, it contains some highly reddened starbursts which do not show the 220-nm feature. Addressing this point would therefore be an important goal for the interpretation of rest-frame UV observations and to get better estimates of the dust attenuation before computing SFRs : we need to analyse a sample of high $S/N$ GALEX spectra containing several galaxy types. 

We list, in Table 1, the SFRs estimated from Kennicutt (1998) relations. After correction for the dust attenuation in FUV estimated from the IR/FUV ratio (Buat et al. 2004), both SFRs correlate with a slope 1.0 and a standard deviation of 6.6 $M_\sun.{\rm yr}^{-1}$ for non-Seyfert galaxies i.e. for those where the IR flux is only coming for recycled star photons and not from a dust torus that would surround the QSO.

\section{Conclusion}

This paper presents the first UV spectra of LIRGs. The analysis of GALEX spectra with complementary $15\mu m$ data from ELAIS allowed us to define a sample of galaxies in the redshift range $0.02 \leq z \leq 0.35$ with multi-wavelength data. We show that the raw GALEX UV color $FUV-NUV$ cannot be used to estimate the UV slope $\beta$ for galaxies at redshift above $z \sim 0.1$ because the GALEX FUV band contains rest-frame flux below $\lambda \sim 1200 \AA$ where spectra flatten. From this multi-$\lambda$ study, we compare two dust attenuation tracers : the UV slope $\beta$ and the IR/UV ratio. We show that for IR-bright objects with an average dust attenuation in FUV : $A_{FUV} = 2.7 \pm 0.8$, the former does not seem to be valid. However, the best S/N galaxy at $z = 0.048$ could show a hint of the presence of a bump approximatively at the same location that the $2175 \AA$ feature in the Milky Way attenuation curve. If we try to correct for the effect of this bump by not using pixels in the range $2000 \AA - 2500 \AA$, the galaxy gets closer to the location of the starburst relationship in the $IR/FUV$ vs. $\beta$ diagram. $S/N$ and/or available wavelength range prevent us from carrying out the same analysis on the rest of the sample and it is clear that we need to observe other high $S/N$, low-redshift galaxies with GALEX.
\acknowledgments

GALEX (Galaxy Evolution Explorer) is a NASA Small Explorer, launched in April 2003.
We gratefully acknowledge NASA's support for construction, operation,
and science analysis for the GALEX mission,
developed in cooperation with the Centre National d'Etudes Spatiales
of France and the Korean Ministry of
Science and Technology. We also thank the French Programme National Galaxies and the Programme National Cosmologie for their financial support.




\clearpage



\begin{figure}
\epsscale{1.}
\plotone{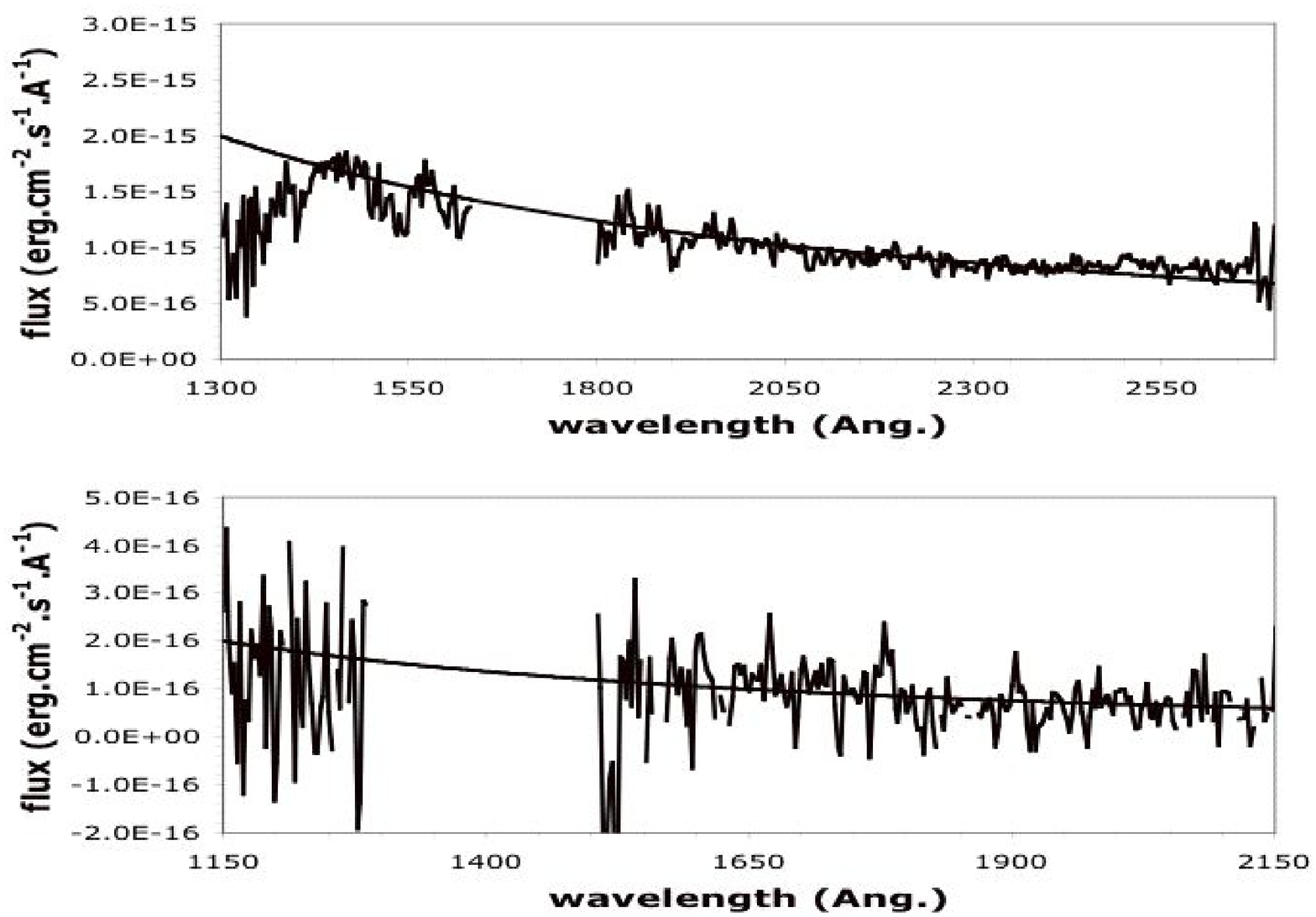}
\caption{\label{fig1} The upper paner shows one of the best fits of a power law on the spectrum of ELAISC15\_J003828-433848 at $z=0.048$ ($\beta=-1.58 \pm 0.04$) and one of the worst for ELAISC15\_J003531-434448 at $z=0.286$ ($\beta=-2.04 \pm 0.27$). Note that because of its redshift, the fit of the latter spectrum was performed on less pixels since we do not use pixels at $\lambda \leq 1200 \AA$.}
\end{figure}

\clearpage\begin{figure}
\epsscale{1.}
\plotone{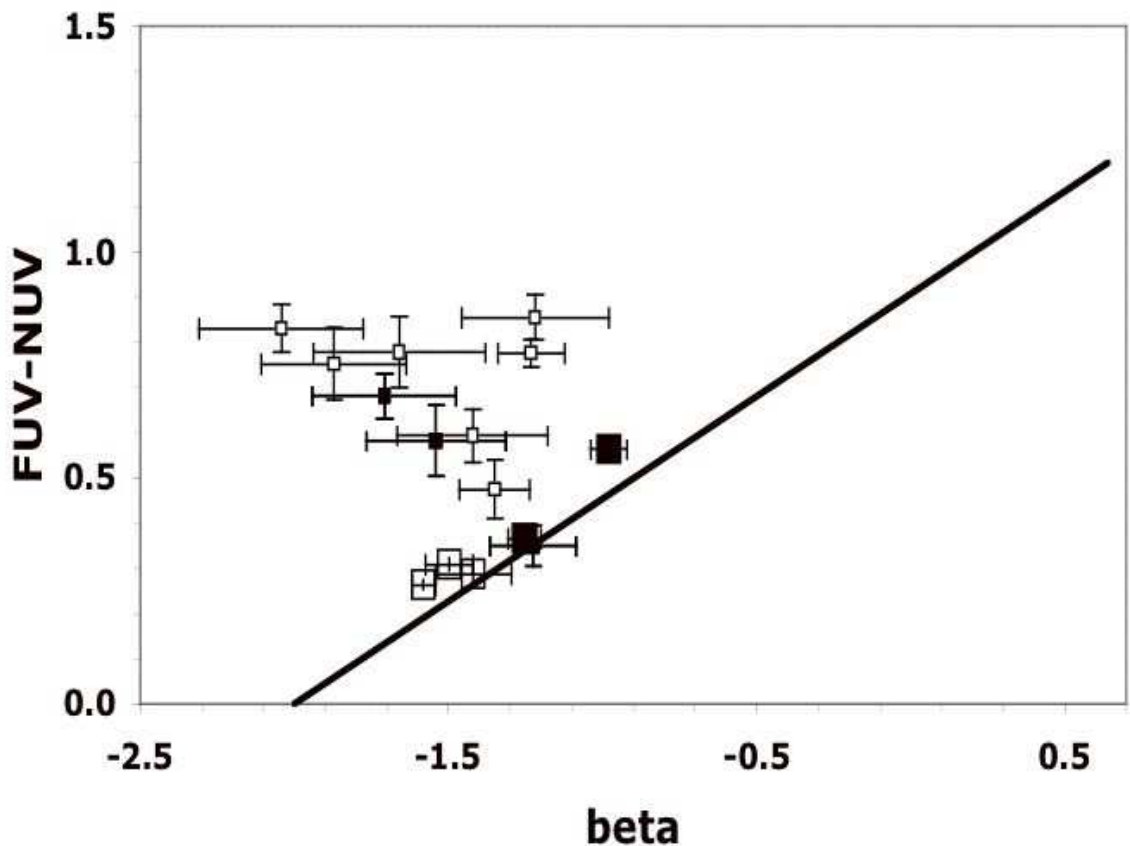}
\caption{\label{fig2} The GALEX FUV-NUV color is plotted against the UV slope $\beta$ estimated from GALEX spectra. Our sample is represented as follows : open boxes are normal galaxies and filled boxes are Sy2 galaxies. We use larger boxes for galaxies with $z < 0.1$. There is no clear relationship between $\beta$ and $FUV-NUV$ for the whole galaxy sample. However, there is a tight relationship if we use only galaxy at $z < 0.1$. Within the GALEX maximal 0.1-0.15 uncertainty, all the lowest redshift galaxies closely follows the law (continuous line) given by Kong et al. (2004) and Kinney et al.'s (1993) templates (crosses) integrated into GALEX bandpasses.}
\end{figure}
\clearpage

\clearpage\begin{figure}
\epsscale{1.}
\plotone{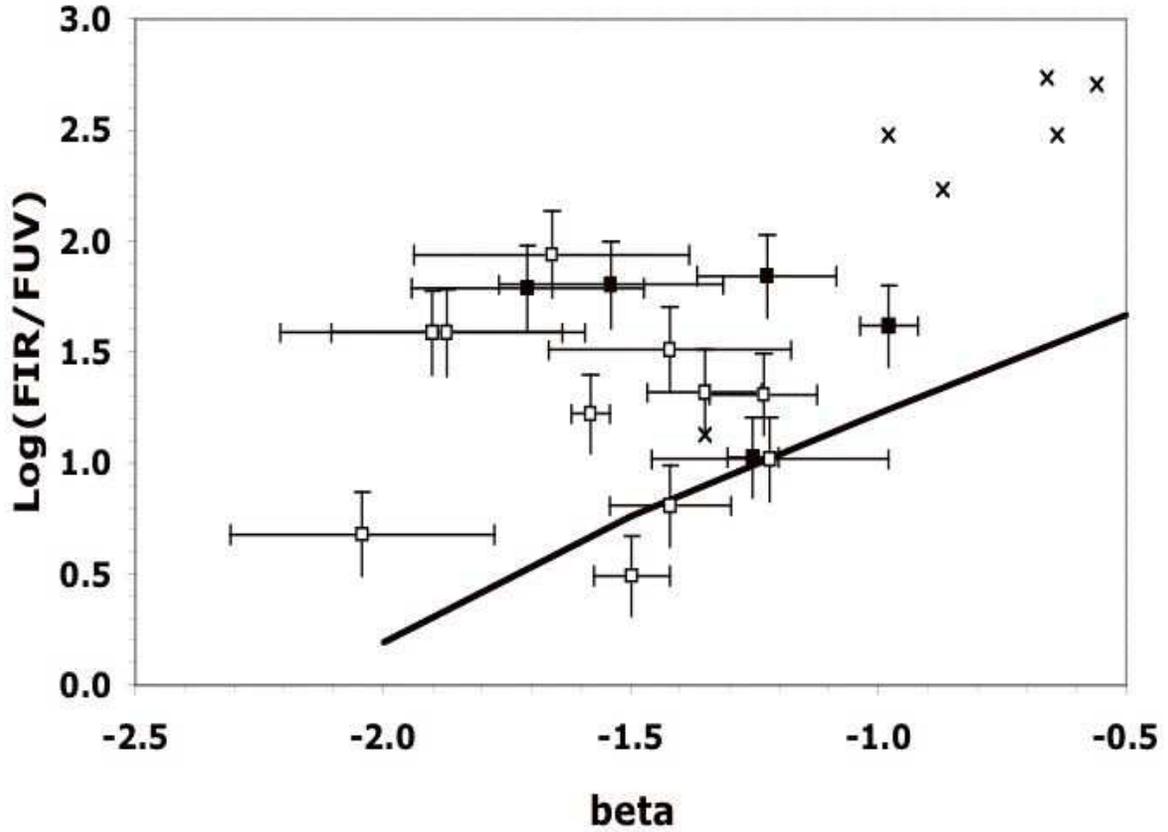}
\caption{\label{fig3} The Log of the L$_{IR}$ / L$_{FUV}$ ratio is plotted against the UV slope $\beta$. Our sample is represented with the same symbols than in Fig. 2 and Goldader et al. (2002) objects are added as crosses. One of them falls within our sample and the other ones are redder and more extinguished. Most of the objects do not follow the law deduced from UV observation of starbursts (e.g. Kong et al. 2004 overplotted here) : the sample extends upwards.}
\end{figure}
\clearpage

\begin{figure}
\epsscale{1.}
\plotone{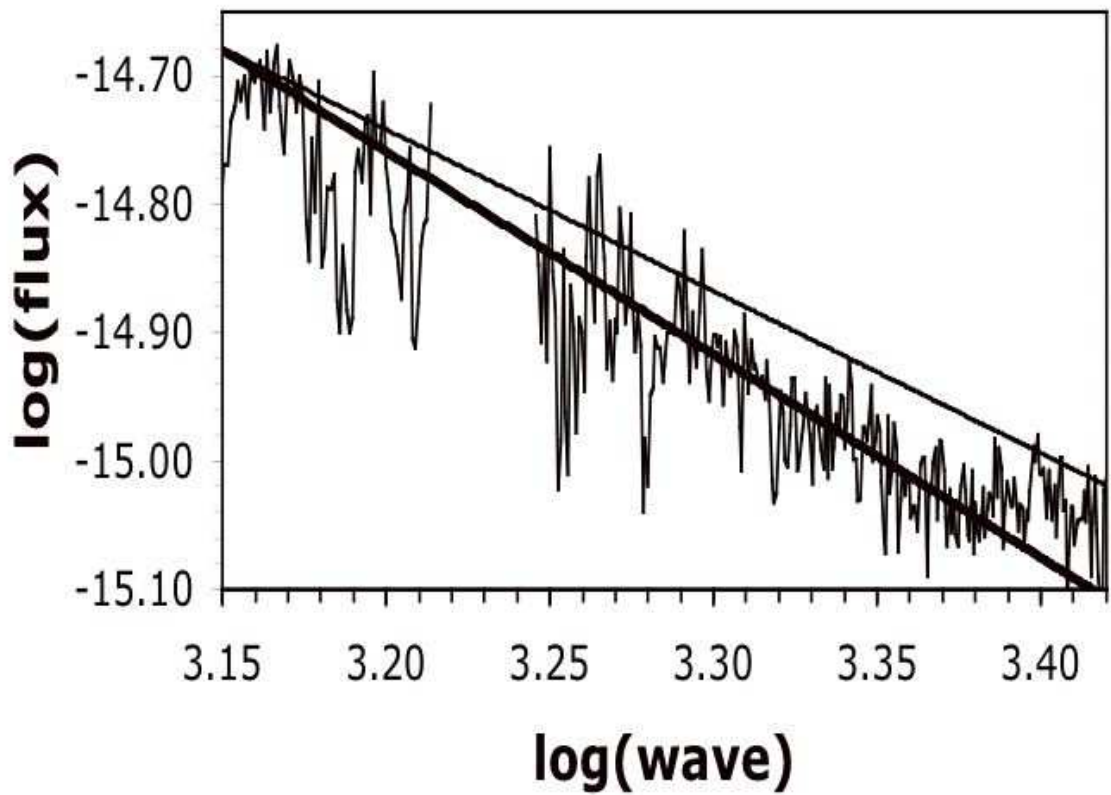}
\caption{\label{fig4} The highest-quality spectra is shown with its fits ($f_\lambda \propto \lambda^\beta$). This object seems to exhibit a trough consistent with being due to a bump around 220 nm ($log_{10}(\lambda) = 3.34)$ similar to the Milky Way Extinction law. The regular fit (heavy line) gives $\beta$ = -1.58 $\pm 0.04$. If we estimate $\beta$ without the trough by excluding pixels in the range $1900 \AA - 2500 \AA$, i.e. $3.28 \leq log_{10} (\lambda) \leq 3.39$ (thin line), we obtain $\beta = -1.26$ at 8 $\sigma$ from the previous value. Units are in erg.s$^{-1}$.cm$^{-2}$.$\AA^{-1}$ and Log($\lambda$) in $\AA$.}
\end{figure}
\clearpage






\clearpage

\begin{deluxetable}{cccccccccc}
\tabletypesize{\scriptsize}
\tablecaption{In this Table, magnitudes are in the AB system. UV slopes $\beta$ are estimated from the available rest-frame range (i.e. $\lambda \lambda 1200 \AA - 3000 \AA$) except for ELAISC15\_J003526-435640 which is at the edge of the field of view and the two last objects for which this rest-frame range is not observed by GALEX because of the redshift. Magnitudes are in the AB system ($m_{AB} = -2.5 Log (f_\nu$ [erg.cm$^{-2}$.s$^{-1}$.Hz$^{-1}$]) - 48.6. Luminosities are in erg.s$^{-1}$. The star formation rates are estimated from Kennicutt (1988) and $A_{FUV}$ from Buat et al. (2002).}
\tablewidth{0pt}
\tablehead{
\colhead{Name} & \colhead{FUV\tablenotemark{1}} & \colhead{NUV\tablenotemark{2}} & \colhead{$\beta (\sigma_\beta)$} & \colhead{$Log(L_{FUV})$} & \colhead{$Log(L_{IR})$\tablenotemark{3}} & \colhead{$z_{spec}$\tablenotemark{4}} & 
\colhead{$A_{FUV}$} & \colhead{SFR$_{UV}$} & \colhead{SFR$_{IR}$}
}
\startdata
ELAISC15\_J003828-433848 & 17.7 & 17.4 & -1.58(0.04) &  9.9 & 11.1 &  0.048 & 2.43 & 19.3 & 19.6 \\
ELAISC15\_J003731-440812 & 18.7 & 18.4 & -1.25(0.05) &  9.6 & 10.6 &  0.052\tablenotemark{5} & 2.06 & &  \\
ELAISC15\_J003645-440720 & 18.4 & 18.1 & -1.50(0.08) &  9.8 & 10.3 &  0.059 & 1.17 & 5.5 & 3.8  \\
ELAISC15\_J003926-441140 & 20.0 & 19.5 & -0.98(0.06) &  9.5 & 11.2 &  0.088\tablenotemark{5} & 3.22 & &  \\
ELAISC15\_J003859-433936 & 20.0 & 19.7 & -1.42(0.12) &  9.8 & 10.6 &  0.119 & 1.67 & 8.4 & 9.2 \\
ELAISC15\_J003938-433755 & 21.9 & 21.3 & -1.54(0.23) &  9.1 & 10.9 &  0.125\tablenotemark{5} & 3.61 & & \\
ELAISC15\_J003530-435604 & 22.0 & 21.2 & -1.66(0.28) &  9.2 & 11.2 &  0.147 & 3.90 & 16.9 & 25.3 \\
ELAISC15\_J003603-435602 & 21.4 & 20.8 & -1.42(0.24) &  9.5 & 11.0 &  0.148 & 2.99 & 12.5 & 4.5 \\
ELAISC15\_J003942-435403 & 21.7 & 21.2 & -1.35(0.12) &  9.4 & 10.7 &  0.149 & 2.61 & 6.8 & 9.4  \\
ELAISC15\_J003932-441130 & 20.2 & 19.5 & -1.23(0.11) & 10.2 & 11.5 &  0.185 & 2.59 & 43.9 & 43.9 \\
ELAISC15\_J003921-441134 & 22.0 & 21.2 & -1.87(0.23) &  9.5 & 11.1 &  0.189 & 3.15 & 15.1 & 19.6 \\
ELAISC15\_J003754-441106 & 21.1 & 20.4 & -1.71(0.23) & 10.0 & 11.7 &  0.212 & 3.58 & 46.0 & 38.8 \\
ELAISC15\_J003721-434239 & 20.9 & 20.5 & -1.23(0.14) & 10.1 & 11.9 &  0.225\tablenotemark{5} & 3.69 & & \\
ELAISC15\_J003716-434153 & 21.2 & 20.3 & -1.22(0.24) & 10.0 & 11.0 &  0.226 & 2.04 & 14.8 & 15.9  \\
ELAISC15\_J003531-434448 & 20.9 & 20.1 & -2.04(0.27) & 10.3 & 11.0 &  0.286 & 1.46 & 25.5 & 14.9 \\
ELAISC15\_J003526-435640 & 21.9 & 21.2 &             & 10.0 & 11.5 &  0.324 & 2.91 & 44.0 & 57.9  \\
ELAISC15\_J003954-440510 & 21.4 & 21.1 & -1.90(0.31) & 10.3 & 11.8 &  0.331 & 3.15 & & \\
ELAISC15\_J003813-433315 & 21.0 & 19.6 &             & 12.0 & 12.9 &  1.400\tablenotemark{6} & & & \\
ELAISC15\_J003829-434454 &      & 19.3 &             &      & 13.8 &  1.567\tablenotemark{6} & & & \\
\enddata



\tablenotetext{1}{Relative error in FUV amounts to about 0.10}
\tablenotetext{2}{Relative error in NUV amounts to about 0.05}
\tablenotetext{3}{Bolometric IR luminosities are computed following Chary \& Elbaz (2002).Absolute errors should be about 0.5 in $Log(L_{IR})$}
\tablenotetext{4}{Spectroscopic redshifts from ELAIS database}
\tablenotetext{5}{Seyfert 2 galaxies from ELAIS database}
\tablenotetext{6}{Seyfert 1 galaxies from ELAIS database}
\end{deluxetable}

\end{document}